\documentclass[prl,twocolumn,showpacs,preprintnumbers,amsmath,amssymb]{revtex4}
\usepackage{graphicx}% Include figure files
\usepackage{dcolumn}% Align table columns on decimal point
\usepackage{bm}% bold math
\usepackage{psfig}
\newcommand \be{\begin{eqnarray}}
\newcommand \ee{\end{eqnarray}}

\begin{document}
%\twocolumn[\hsize\textwidth\columnwidth\hsize
%           \csname @twocolumnfalse\endcsname
\title{Fluctuations due to the nonlocal character of collisions}
\author{K. Morawetz$^{1,2}$
}
\affiliation{$^1$Institute of Physics, Chemnitz University of Technology, 
09107 Chemnitz, Germany}
\affiliation{
$^2$Max-Planck-Institute for the Physics of Complex
Systems, N{\"o}thnitzer Str. 38, 01187 Dresden, Germany}
\begin{abstract}
It is shown that the collision integral describing
the nonlocal character of collisions leads to the same mean-field
fluctuations as proposed by Boltzmann-Langevin pictures. It is argued
that this appropriate collision integral contains the 
fluctuation-dissipation theorems in equilibrium itself and 
therefore there is no need to assume additionally stochasticity.
This leads to tremendous simplifications in numerical simulation schemes.
\end{abstract}
%\vskip2pc]
\maketitle

The question how to describe sufficiently the fluctuations in dynamical 
systems of many interacting particles is as old as the discovery of 
the Brownian motion. Different schemes have been developed in different
branches of physics. Mostly additional stochasticity is assumed to account 
for such fluctuations. The goal of this paper is to show how the correct 
fluctuations can be described by a realistic collision scenario within 
the deterministic kinetic theory without ad-hoc assumptions about 
stochasticity. 
Here we will briefly outline the attempts in nuclear 
physics having its counterparts in other fields, of course.

It has been noticed in the beginning of the 80th that the time-dependent 
mean-field description (TDHF) of nuclear collisions cannot
describe the experimental fluctuations of observables $Q$ like mass,
charge and momenta of emitted particles \cite{MK85} if the naive
expectation
\be
\Delta Q^2 (t) ={\rm Tr}[Q^2 \hat\rho_n(t)]-[Tr Q \hat\rho_n(t)]^2
\label{2}
\ee
with the TDHF density matrix $\hat\rho_n$ is applied. This deficiency has been
cured by a variational approach \cite{BV81,TV85,T85,T86,Fl89} which
leads to the propagation of fluctuations from $t_0$ 
\be
\Delta Q^2(t_1)=\lim\limits_{\eta \to 0} {1 \over 2
  \eta^2}{\rm Tr}[\hat \rho_n(t_0,0)-\hat\rho_n(t_0,\eta)]^2
\label{1}
\ee
where $\hat\rho_n(t,\eta)$ evolves with the TDHF equation but with the boundary
condition
%\be
$\hat\rho_n(t,\eta)=\exp{(-i \eta Q)}\hat\rho_n(t)\exp{(i \eta Q)}$.
%\ee
In this way the fluctuations are obtained by propagating back in time
but they become explicitly dependent on the observables $Q$. It has been
shown that the expression (\ref{1}) leads besides the TDHF fluctuation
(\ref{2}) to an additional part which can be described by higher order
diagrams in the interaction \cite{T85}. The application of this
procedure leads to a significant enhancement of the fluctuations
\cite{MK85,ZG88,LCV89}. 
The fact that higher
order diagrams are necessary to describe more appropriate
fluctuations shows that the
collisions are not described appropriately in usual Boltzmann (BUU) 
simulations.

Alternatively there has been developed the
time dependent generator coordinator method (TDGCM) \cite{RGC80,RS92} 
which expands
the wave function $\Psi$ in a set of TDHF wave functions $\phi_N$
\be
|\Psi(t)\rangle =\sum\limits_N|\phi_N(t)\rangle  f_N(t)
\ee
where the coefficients $f_N$ are determined from minimizing the
action.
The TDGCM wave function leads to optimal fluctuations if the TDHF
basic set $\phi_N(0)$ is chosen such that the variable $Q$ becomes a
generator of the path at finite time $t$.
%\be
%|\phi_N(0)\rangle =U_{MF}(0,t) {\rm e}^{i Q t} U_{MF}(t,0)|\phi_0(0)\rangle .
%\ee 
This TDGCM schema is equivalent to the above described 
Balian-Veneroni variational
approach \cite{BV81} in the random phase approximation (RPA) limit. 
The advantage of the TDGCM schema is that
it provides for optimal paths. 
However, both schemes are too limited for practical applications since
one can handle only small sets of collective correlation channels.

This practical limitation has led to the development of stochastic
TDHF \cite{RS92} which approximates the time evolution of the N-particle density
operator ${\hat \rho}$ at a small time steps $t_i$ by the diagonal elements of the expansion in TDHF density
operators ${\hat \rho}_n$
\be
{\hat \rho}(t_i)=\sum\limits_n W_n {\hat \rho}_n(t_i)+\sum\limits_{nn'} W_n P_{n'n} t_i
[{\hat \rho}_{n'}(t_i)-{\hat \rho}_n(t_i)].
\label{Stoch}
\ee
The transition probability is given by the matrix element of
TDHF Slater states at time $t_i$
\be
P_{nn'}={2 \pi \over \hbar}|\langle n|V|n'\rangle |^2 \delta(E_n-E_n')
\ee
which leads to a Monte Carlo method of evolving an initial state into
an ensemble of Slater states with the probability 
$
W_m=1-\sum\limits_{n'} P_{n'n} t_i$ for $m=n$ and $W_m=
P_{mn}$ for $m\ne n$.

The usually used
Boltzmann collision terms with the inclusion of Pauli-blocking (BUU)
cannot account for these fluctuations since the collisions are treated
as ideal, i.e. local in space and time. Therefore there has been proposed
another method of including more realistic fluctuations in the Boltzmann 
equation for the one-particle distribution $f$ 
by adding a stochastic term $\delta I$ to the collision
integral, called Boltzmann - Langevin picture 
\cite{AG90,SASB90,RR90,RSA92,ASBB92,BGS92a,BCR91}
\be
{d f\over dt}=(1-f) W_{\rm in}-f W_{\rm out} +\delta I
\label{3}
\ee
where schematically the scattering - out and -in probability of a
phase space cell is $W_{\rm out/in}$. This Boltzmann-Langevin equation can be
formally derived from the stochastic TDHF equation (\ref{Stoch}) if
the one-particle reduced density is traced out.
This equation has been applied for simulation of 
heavy-ion collisions quite frequently
\cite{ColBur93,ColDiT95,CTGMZW98,FCD98,CoCh94,CH96,R98,F01}.
In these treatments the Langevin term in (\ref{3}) is mostly  assumed to be
Markovian
\be 
\langle\delta I\rangle =0, \qquad \langle\delta I(t) \delta I(t')\rangle =2 D \delta (t-t')
\label{i}
\ee
which leads with (\ref{3}) to the equation of motion for the variance
$\sigma^2=\langle f^2\rangle -\langle f\rangle ^2$
\be
{ d \sigma^2 \over dt} =- {2 \over \tau} \sigma^2+ 2 D
\label{4}
\ee
with $1/\tau=W_{\rm out}+W_{\rm in}$. From the stochastic TDHF equation (\ref{Stoch}) 
the form (\ref{i})
follows precisely neglecting fluctuations in the potential i.e. higher
order diagrams \cite{RS92}. Again this is a hint that the
collisions have to be treated more appropriately. Before suggesting a
way to describe fluctuations more realistically, let us discuss some
principle problems of Boltzmann-Langevin approaches. 

We have in principle no reason to see the time evolution of the density 
operator stochastically since 
the basic van-Neumann equation is deterministic and subsequent 
derived equations should be so as well.
The ad-hoc assumption about stochasticity can mimic the numerical 
noise unavoidable in solving such equations and to a certain extent 
higher-order correlations. Theoretically it is a problem since the
collision integral emerges itself from averaging  about small scale
fluctuations \cite{KL82}. Therefore it is ambiguous to divide correlations
into an averaged collision integral and a stochastic term miming higher-order
correlations. This has been sometimes motivated by the need of
the fluctuation-dissipation theorem associating the collision integral
with dissipation.
In contrast one should observe that the
fluctuation-dissipation theorem emerges itself from appropriate
collision integrals alone since they vanish in equilibrium. This can be
seen best from ring summation approximation (RPA, GW, bubble...) 
leading to the Lenard-Balescu 
collision integral \cite{Moa93,MTM96}. The latter one vanishes exactly
if the quantum fluctuation-dissipation theorem is fulfilled. In other
words if the collision integral is derived appropriately it leads to 
an equilibrium with correct  fluctuation-dissipation theorem and 
there is no need for additional stochastic terms.

Introducing fluctuations the determination of $f$ and the 
stochastic process
$\delta I$ remains phenomenological in the sense that they account
partially for such an appropriate collision integral. This is a practical
need if the appropriate collision integral is
not solvable and one is restricted to Born (Boltzmann) approximation.
Here in this letter we will show that the same mean fluctuations are 
generated from the deterministic but more
realistic nonlocal extension of the Boltzmann collision
integral \cite{SLM96,LSM97,MLSK98,MLSCN98}. The advantage is that besides 
the microscopic foundation the latter one has already established a practical and fast numerical method.

Let us return to the Boltzmann-Langevin equation.
The fluctuation term derived from stochastic TDHF leads to the
educated guess \cite{CTGMZW98}
\be
2 D=(1-\langle f\rangle ) W_{\rm in} + \langle f\rangle  W_{\rm out} 
\label{5}
\ee
which has suggested an interesting procedure to include fluctuations
dynamically in BUU codes. From (\ref{4}) one sees that the choice
(\ref{5}) reproduce the equilibrium variance $\sigma^2=f_0 (1-f_0)$
and that the deviation between the actual variance and the statistical
value $\delta \sigma^2 =\sigma^2 -\langle f\rangle (1-\langle f\rangle )$ obeys
\be
{d \delta \sigma^2 \over dt}=- {2 \over \tau} \delta \sigma^2.
\label{6}
\ee
Since $\delta \sigma^2=0$ is a solution of (\ref{6}), 
the averaged value $\langle f\rangle $ can be parametrized as local
equilibrium value such that at a given phase-space cell and time
a small propagation will not deviate the variance
from this result \cite{CTGMZW98}. 
Numerically the authors in \cite{CTGMZW98} used
instead of the variance in distribution the density projection
\be
\sigma_n^2(r,t)=\frac 4 V \int {d p \over \hbar^3}\langle f\rangle (1-\langle f\rangle )
\label{fluc}
\ee  
which has been realized in each phase-space point of collisions.

Besides the practical success of such descriptions in numerical
solutions it remains the more principle question of the validity of the
ad-hoc stochastic assumption. As we have pointed out this assumption
partially cures the ideal collision scenario of space and time
point-like particles. We will follow now the other point of view and
claim that if the collisions are described more realistically by nonlocal
events then the fluctuations should be correctly induced by itself. 

The nonlocal extension of the Boltzmann equation has been given in
\cite{SLM96,LSM97} and the finite duration and dynamical size of
nucleons are calculated in \cite{MLSK98}. The implementation in BUU
codes has allowed to describe experimentally noticeable effects 
\cite{MLSCN98,MTP00,MLNCCT01}. The prediction of a change in the 
reaction mechanisms \cite{MT00} has been nicely confirmed by 
proper scaling of experimental data \cite{BPCBDF01}. The 
correlations by the nonlocal character of collisions are capable to describe long range order as typical for  phase transitions \cite{MTP00}.

These nonlocal extensions are more realistic than the Boltzmann
equation since they lead to the inclusion of two-particle correlations
while in 
the Boltzmann equation these correlations are absent. We will show now that
this more realistic description of the individual collision event
induces also more realistic fluctuations which should be compared to the
above described treatments. Especially we will find
that the nonlocal corrections lead to induced mean-field
fluctuations of the form (\ref{fluc}).

The nonlocal quasiparticle kinetic
equation for the distribution function of particle $a$, $f_1\equiv f_a(k,r,t)$, derived within the non-equilibrium Green's
function technique \cite{SLM96,LSM97} has the form of a Boltzmann equation with the quasiparticle energy $\varepsilon_1=\varepsilon(k,r,t)$
\begin{eqnarray}
\!\!\!\!\!\!\!\!\!\!\!\!\!&&{\partial f_1\over\partial t}+{\partial\varepsilon_1\over\partial k}
{\partial f_1\over\partial r}-{\partial\varepsilon_1\over\partial r}
{\partial f_1\over\partial k}
%\nonumber\\
%\!\!\!\!\!\!\!\!\!\!\!\!\!&&
=s\sum\limits_b\int{dpdq\over(2\pi)^5\hbar^7}{\cal P}_\pm
\nonumber\\
\!\!\!\!\!\!\!\!\!\!\!\!\!&&\times\Bigl[\bigl(1\!-\!f_1\bigr)\bigl(1\!-\!f_2^-\bigr)f_3^-f_4^--
f_1f_2^\pm\bigl(1\!-\!f_3^\pm\bigr)\bigl(1\!-\!f_4^\pm\bigr)\Bigr]
\label{kin}
\end{eqnarray}
and the spin-isospin etc degeneracy $s$.
The superscripts $\pm$ denote the signs of non-local corrections:
$f_2^\pm\equiv f_b(p,r\!\pm\!\Delta_2,t)$,
$f_3^\pm\equiv f_a(k\!-\!q\!\pm\!\Delta_K,r\!\pm\!\Delta_3,t\!\pm\!
\Delta_t)$, and $f_4^\pm\equiv f_b(p\!+\!q\!\pm\!\Delta_K,r\!\pm\!
\Delta_4,t\!\pm\!\Delta_t)$. For the out-scattering part of
(\ref{kin}) both signs can be given equivalently \cite{SLM98}.  
The scattering measure is given by the
modulus of the scattering T-matrix ${\cal P}_\pm=|{\cal T}^R_\pm|^2 \delta (\varepsilon_1+\varepsilon_2-\varepsilon_3-\varepsilon_4\pm 2 \Delta_E)$.
All corrections,
the $\Delta$'s, describing the non-local and non-instant collision are given by derivatives of the scattering phase shift
\mbox{$\phi={\rm Im\ ln}{\cal T}^R(\Omega,k,p,q,t,r)$}
\begin{equation}
\begin{array}{lclrcl}\Delta_t&=&{\displaystyle
\left.{\partial\phi\over\partial\Omega}
\right|_{\varepsilon_1+\varepsilon_2}}&\ \ \Delta_2&=&
{\displaystyle\left({\partial\phi\over\partial p}-
{\partial\phi\over\partial q}-{\partial\phi\over\partial k}
\right)_{\varepsilon_1+\varepsilon_2}}\\ &&&&&\\ \Delta_E&=&
{\displaystyle\left.-{1\over 2}{\partial\phi\over\partial t}
\right|_{\varepsilon_1+\varepsilon_2}}&\Delta_3&=&
{\displaystyle\left.-{\partial\phi\over\partial k}
\right|_{\varepsilon_1+\varepsilon_2}}\\ &&&&&\\ \Delta_K&=&
{\displaystyle\left.{1\over 2}{\partial\phi\over\partial r}
\right|_{\varepsilon_1+\varepsilon_2}}&\Delta_4&=&
{\displaystyle-\left({\partial\phi\over\partial k}+
{\partial\phi\over\partial q}\right)_{\varepsilon_1+\varepsilon_2}}.
\end{array}
\label{SHIFTSALL}
\end{equation}
The nonlocal kinetic equation (\ref{kin}) covers all quantum virial corrections on the
binary level and conserves density, momentum and energy including the
corresponding two-particle correlated parts \cite{LSM97}. It requires
no more computational power than solving the Boltzmann equation \cite{MLSCN98,MLNCCT01}.

We will now derive the fluctuation contribution to the mean-field from
this collision integral. Therefore we use the fact that the mean-field
can be considered as zero-angle collisions, a collision which
does not change momenta of particles but changes its phase. Summing up
all these changes in phase gives just the mean-field potential due to
the surrounding media. When deriving collision integrals
one tries to share correlations in such a way that 
all mean-field like contributions are collected in the
quasiparticle energies on the left side and all true collisions with
finite transferred momenta are on the right side. In addition we will 
observe
now that due to the nonlocal character of the collision there is a
finite zero transfer momenta channel hidden in the nonlocal collision
integral which can be rewritten into the drift side of the kinetic
equation and which gives exactly the fluctuations (\ref{fluc}).

To show this we rewrite the energy-conserving $\delta$-function  in 
(\ref{kin}) as 
\be
&&\delta\left ({k^2\over 2 m_a}+{p^2\over 2 m_b}-{(k-q)^2\over 2 m_a}-{(p+q)^2\over 2 m_b}\right )\nonumber\\
&&=\delta\left (|q|({k\over m_a}-{p\over m_b})\cdot {q\over
    |q|}-{q^2\over 2}
  ({1\over m_a} + {1\over m_b})\right )\nonumber\\
&&={\delta(|q|) \over \left ({k\over m_a}-{p\over m_b}\right)\cdot
    {q\over |q|}} + \delta(q\ne0)
\label{d}
\ee
where the channel $q\ne0$
represents  the usually collision integral of (\ref{kin}).
The
$q=0$ channel leads now to an additional part absent in usual local kinetic
equations like the Boltzmann equation. 
To convince the reader about this novel observation let us rewrite the
Pauli-blocking factors of  (\ref{kin}) for the $q=0$ channel according
to (\ref{d})
\be
&&\Bigl[\bigl(1\!-\!f_1\bigr)\bigl(1\!-\!f_2^-\bigr)f_3^-f_4^--
f_1f_2^\pm\bigl(1\!-\!f_3^\pm\bigr)\bigl(1\!-\!f_4^\pm\bigr)\Bigr]_{q=0}
\nonumber\\
&&=f_2^-(1-f_2^-) \Bigl[f_1^--f_1 \Bigr].
\ee
We see that in the case of local kinetic equations without delays
this specific channel disappears since $f_1^-=f_1$. Therefore in
contrast to the usual local kinetic equations the nonlocal 
equation possesses a finite zero-angle channel in the collision
integral which is of mean-field type since no energy or
momenta is exchanged. The fact that retardation leads to an additional
correction to the Bogoliubov--Hartree-Fock mean-field has been
observed first within linear response in \cite{GRWT93}.

From (\ref{kin}) we obtain now
\begin{eqnarray}
\!\!\!\!\!\!\!\!\!\!\!\!\!
&&{\partial f_1\over\partial t}\!+\!{\partial\varepsilon_1\over\partial k}
{\partial f_1\over\partial r}\!-\!{\partial\varepsilon_1\over\partial r}
{\partial f_1\over\partial k}={\langle f_1^-\!-\!f_1\rangle  \over \tilde \tau}
%\nonumber\\
%\!\!\!\!\!\!\!\!\!\!\!\!\!&&
\!+\!s\!\!\int\limits_{q\ne0}\!\!{dpdq\over(2\pi)^5\hbar^7}{\cal P}_\pm
\nonumber\\
\!\!\!\!\!\!\!\!\!\!\!\!\!
&&\times\Bigl[\bigl(1\!-\!f_1\bigr)\bigl(1\!-\!f_2^-\bigr)f_3^-f_4^--
f_1f_2^\pm\bigl(1\!-\!f_3^\pm\bigr)\bigl(1\!-\!f_4^\pm\bigr)\Bigr]
\label{kin1}
\end{eqnarray}
where
\be
\!\!{1\over \tilde \tau}\!=\!\big\langle {2 \pi s \over \hbar} \!\!\!\int \!\!{d p dq \over (2 \pi
  \hbar)^6}
|{\cal T}^r|^2 
%\nonumber\\&& \times
\delta\!\left (\!{k\cdot q\over m_a}-{p\cdot q\over m_b}\!\right ) 
f_p(1-f_p)\big\rangle .
\label{tt}
\ee
Here we use the approximation of thermal averaged delays, $\langle ...\rangle $, which allows
us to pull out the $f^-_1$ term which contains the $p$-dependent shift
under the integration of the $1/\tilde\tau$ term.  This serves here for legibility and can
be rendered exactly if one keeps the corresponding shift terms under
the $(p,q)$ integral of (\ref{tt}). 

The resulting equation (\ref{kin1}) is
a delay-differential equation and has lead already 
to an interesting interplay between 
stochastic bifurcations and relaxation due to inertia \cite{Mc01}. We see 
from (\ref{kin1}) and (\ref{tt}) that terms 
$1/\tilde\tau \propto \sigma^2$ of
(\ref{fluc}) appears. In order to see that this additional term
represents mean-field fluctuations let us expand
\be
\!\!\!\!\!\!\langle f_1\!-\!f_1^-\rangle \!=\!\langle \Delta_t \!{\partial f \over \partial t}\!+\!\Delta_3 \!{\partial f
  \over \partial r}\!+\!\Delta_K \!{\partial f \over \partial k}\rangle 
%\nonumber\\
%&=&
\!\!\approx\!\!\tilde\Delta_3 \!{\partial f
  \over \partial r}\!+\!\tilde\Delta_K \!{\partial f \over \partial k}
\label{form}
\ee
where in the last step we have replace the time derivative of $f$ by
the free drift motion ${\partial f_1\over\partial t}\approx -{\partial\varepsilon_1\over\partial k}
{\partial f_1\over\partial r}+{\partial\varepsilon_1\over\partial r}
{\partial f_1\over\partial k}$ leading to the on-shell shifts
\be
\tilde \Delta_3&=&\langle \Delta_3-\Delta_t {\partial\varepsilon_k\over\partial
  k}\rangle =
-\langle {\partial\phi^{on}\over\partial k}\rangle \nonumber\\
\tilde \Delta_K&=&\langle \Delta_K+\Delta_t {\partial\varepsilon_k\over\partial
  r}\rangle =
\langle {\partial\phi^{on}\over\partial r}\rangle .
\label{on}
\ee
Now we can shift (\ref{form}) from the right to the left (drift) side of
  (\ref{kin1})
to obtain finally
\begin{eqnarray}
&&\!\!\!\!\!\!\!\!\!\!\!\!\!{\partial f_1\over\partial t}+\left ({\partial\varepsilon_1\over\partial
  k}+{\tilde\Delta_3\over \tilde \tau}\right )
{\partial f_1\over\partial r}-\left ({\partial\varepsilon_1\over\partial r}-{\tilde\Delta_K\over \tilde \tau}\right )
{\partial f_1\over\partial k}
\nonumber\\
&&\!\!\!\!\!\!\!\!\!\!\!\!\!=s\sum\limits_b\int\limits_{q\ne0}{dpdq\over(2\pi)^5}{\cal P}_\pm
\nonumber\\
&&\!\!\!\!\!\!\!\!\!\!\!\!\!\times\Bigl[\bigl(1\!-\!f_1\bigr)\bigl(1\!-\!f_2^-\bigr)f_3^-f_4^--
f_1f_2^\pm\bigl(1\!-\!f_3^\pm\bigr)\bigl(1\!-\!f_4^\pm\bigr)\Bigr].
\label{kin2}
\end{eqnarray}
We see that an explicit fluctuating term 
$\sim 1/\tilde\tau \sim \sigma_n^2(r,t)$ emerges to the mean
field which has the form of (\ref{fluc}). 

Therefore we conclude that the variance of fluctuations from the
Langevin--Boltzmann equation can be reproduced in a deterministic way 
from the nonlocal 
extension of the Boltzmann equation. The stochastic treatment in
numerical realizations can be considered therefore as a numerical trick to 
reproduce the correct deterministic fluctuations. The nonlocal collision scheme instead provides a first principle theory of such fluctuations.
It gives probably
the same practical results but with much less numerical effort since the collision scenario of the usual  BUU code is modified only slightly with no additional computational time required \cite{MLSCN98,MLNCCT01}. Therefore we have demonstrated that the ad-hoc assumption about 
Langevin sources to the Boltzmann equation is unnecessary if the 
collisions are treated non-locally.

The clarifying discussions with P.G. Reinhard are gratefully acknowledged.

\bibliography{kmsr,kmsr1,kmsr2,kmsr3,kmsr4,kmsr5,kmsr6,kmsr7,delay2,spin,refer,delay3,gdr,chaos,sem3,sem1,sem2,short}
\end{document}